\documentclass[aps,prl,superscriptaddress,preprintnumbers,twocolumn,groupedaddress,showpacs]{revtex4}

\usepackage{graphicx}

\newcommand{\rT}{{\mathrm{T}}}
\newcommand{\ri}{{\mathrm{i}}}
\newcommand{\weak}{{\mathrm{weak}}}
\newcommand{\LO}{{\mathrm{LO}}}

\newcommand{\EW}{{\mathrm{EW}}}

\newcommand{\ppmmee}{\Pp \Pp \to \mu^+\mu^-\Pe^+\Pe^- + X}
\newcommand{\lsim}
{\mathrel{\raisebox{-.3em}{$\stackrel{\displaystyle <}{\sim}$}}}

\def\mathswitchr#1{\relax\ifmmode{\mathrm{#1}}\else$\mathrm{#1}$\fi}
\newcommand{\PZ}{\mathswitchr Z}
\newcommand{\PH}{\mathswitchr H}

\newcommand{\Pe}{\mathswitchr e}
\newcommand{\Pep}{\mathswitchr {e^+}}
\newcommand{\Pem}{\mathswitchr {e^-}}
\newcommand{\Pp}{\mathswitchr p}
\newcommand{\Pg}{\mathswitchr g}
\newcommand{\Pt}{\mathswitchr t}

\newcommand{\PW}{\mathswitchr W}

\def\mathswitch#1{\relax\ifmmode#1\else$#1$\fi}

\newcommand{\MW}{\mathswitch {M_\PW}}

\newcommand{\MZ}{\mathswitch {M_\PZ}}
\newcommand{\MH}{\mathswitch {M_\PH}}

\newcommand{\Mt}{\mathswitch {m_\Pt}}
\newcommand{\GW}{\mathswitch {\Gamma_\PW}}
\newcommand{\GZ}{\mathswitch {\Gamma_\PZ}}

\newcommand{\TeV}{\unskip\,\mathrm{TeV}}
\newcommand{\GeV}{\unskip\,\mathrm{GeV}}

\def\refeq#1{\mbox{(\ref{#1})}}

\def\reffi#1{\mbox{Fig.~\ref{#1}}}

\def\citere#1{\mbox{Ref.~\cite{#1}}}
\def\citeres#1{\mbox{Refs.~\cite{#1}}}

\def\draftdate{\relax}
\def\mda{\relax}
\def\mua{\relax}
\def\mla{\relax}
\def\draft{
\def\thtystars{******************************}
\def\sixtystars{\thtystars\thtystars}
\typeout{}
\typeout{\sixtystars**}
\typeout{* Draft mode!
         For final version remove \protect\draft\space in source file *}
\typeout{\sixtystars**}
\typeout{}
\def\draftdate{\today}
\def\mua{\marginpar[\boldmath\hfill$\uparrow$]%
                   {\boldmath$\uparrow$\hfill}%
                    \typeout{marginpar: $\uparrow$}\ignorespaces}
\def\mda{\marginpar[\boldmath\hfill$\downarrow$]%
                   {\boldmath$\downarrow$\hfill}%
                    \typeout{marginpar: $\downarrow$}\ignorespaces}
\def\mla{\marginpar[\boldmath\hfill$\rightarrow$]%
                   {\boldmath$\leftarrow$\hfill}%
                    \typeout{marginpar: $\leftrightarrow$}\ignorespaces}
\def\Mua{\marginpar[\boldmath\hfill$\Uparrow$]%
                   {\boldmath$\Uparrow$\hfill}%
                    \typeout{marginpar: $\uparrow$}\ignorespaces}
\def\Mda{\marginpar[\boldmath\hfill$\Downarrow$]%
                   {\boldmath$\Downarrow$\hfill}%
                    \typeout{marginpar: $\downarrow$}\ignorespaces}
\def\Mla{\marginpar[\boldmath\hfill$\Rightarrow$]%
                   {\boldmath$\Leftarrow $\hfill}%
                    \typeout{marginpar: $\leftrightarrow$}\ignorespaces}
\overfullrule 5pt
\oddsidemargin -10mm
\marginparwidth 15mm
}

\begin{document}

\preprint{FR-PHENO-2016-002, ICCUB-16-003}
\title{\boldmath{
Electroweak corrections to $\Pp\Pp\to\mu^+\mu^-\Pep\Pem+X$ at the LHC \\
--- \\
a Higgs background study}}
 
\author{B.~Biedermann}
\affiliation{Julius-Maximilians-Universit\"at W\"urzburg,
Institut f\"ur Theoretische Physik und Astrophysik, 
D-97074 W\"urzburg, Germany}

\author{A.~Denner}
\affiliation{Julius-Maximilians-Universit\"at W\"urzburg,
Institut f\"ur Theoretische Physik und Astrophysik, 
D-97074 W\"urzburg, Germany}

\author{S.~Dittmaier}
\affiliation{Albert-Ludwigs-Universit\"at Freiburg, Physikalisches Institut, 
D-79104 Freiburg, Germany}

\author{L.~Hofer}
\affiliation{Department d'Estructura i Constituents de la Mat\`eria (ECM),
Institut de Ci\`encies del Cosmos (ICCUB),
Universitat de Barcelona (UB),
Mart\'i Franqu\`es 1,
E-08028 Barcelona, Spain}

\author{B.~J\"ager}
\affiliation{Eberhard-Karls-Universit\"at T\"ubingen, Institut f\"ur
Theoretische Physik, D-72076 T\"ubingen, Germany}

\date{\today}

\begin{abstract}
  The first complete calculation of the next-to-leading-order
  electroweak corrections to four-lepton production at the LHC is
  presented, where all off-shell effects of intermediate Z~bosons and 
photons are taken into account.  Focusing on the mixed final state
$\mu^+\mu^-\Pep\Pem$, we study differential cross sections that are
particularly interesting for Higgs-boson analyses. The electroweak
corrections are divided into photonic and purely weak corrections.
The former exhibit patterns familiar from similar W/Z-boson production
processes with very
large radiative tails near resonances and kinematical shoulders.  The
weak corrections are of the generic size of $5\%$ and show
interesting variations, in particular a sign change between the
regions of resonant Z-pair production and the Higgs signal.
\end{abstract}

\pacs{12.15.Ji, 12.15.Lk}

\maketitle

\subsection{Introduction}

The investigation of pair production processes of electroweak (EW)
gauge bosons W, Z, and $\gamma$ is of great importance at the CERN
Large Hadron Collider (LHC).  These processes have sizeable cross
sections and provide experimentally clean signatures via the leptonic
decay modes of the W or Z~bosons.  On the one hand, they offer an
indirect window to potential new-physics effects through their
sensitivity to the self-interactions among the EW gauge bosons; on the
other hand, these reactions represent sources of irreducible
background to many direct searches for new particles (e.g.\ additional
heavy gauge bosons $\PW^\prime,\PZ^\prime$) and to precision studies
of the Higgs boson discovered in 2012 in particular.

In order to optimally exploit and interpret LHC data, theoretical
predictions to weak-gauge-boson pair production have to be pushed to
an accuracy at the level of percent, a task that requires the
inclusion of higher-order corrections of the
strong and EW interactions
and of decay and off-shell effects of the W/Z~bosons. In this
paper we focus on the reaction $\ppmmee$, which does not only include
doubly-resonant ZZ~production, but also interesting regions in phase
space where at least one of the Z~bosons is far off shell, as for
example observed in the important Higgs decay channels
$\PH\to4\,$leptons.

Precision calculations for Z-boson pair production with leptonic
decays have been available for a long time including next-to-leading
order (NLO) QCD
corrections~\cite{Ohnemus:1994ff,Campbell:1999ah,Dixon:1999di}.  They
have even been pushed to next-to-next-to-leading order (NNLO) accuracy
recently~\cite{Cascioli:2014yka,Grazzini:2015hta}, with a
significant contribution from gluon--gluon fusion calculated already
before~\cite{Matsuura:1991pj,Zecher:1994kb,Binoth:2008pr}.  Beyond
fixed perturbative orders, NLO QCD corrections were matched to a
parton shower in
\citeres{Nason:2006hfa,Hoche:2010pf,Hamilton:2010mb,Melia:2011tj,Frederix:2011ss};
in \citere{Cascioli:2013gfa} even different jet multiplicities were
merged at NLO QCD.  Electroweak corrections at NLO are only completely
known for stable Z~bosons~\cite{Bierweiler:2013dja,Baglio:2013toa},
and in some approximation including leptonic decays of on-shell
Z~bosons~\cite{Gieseke:2014gka}.  The EW corrections to Z-pair
production with off-shell Z~bosons, on the other hand, are not yet
known.  In this paper, we fill this gap and present results of the
first full NLO EW calculation for the process $\ppmmee$ in the
Standard Model, including all off-shell contributions.  This allows
us, in particular, to investigate EW corrections in the yet unexplored
kinematic region below the ZZ~threshold, where direct Z-pair
production is an important background to Higgs-boson analyses.

\subsection{General setup of the calculation}

At leading order (LO), the production of $\mu^+\mu^-\Pep\Pem$ final states almost exclusively
proceeds via quark--antiquark annihilation. Contributions from $\gamma\gamma$~collisions
are extremely small 
(they contribute only at the level of a few per mille to the total
cross section) 
owing to the suppression of the photon density in the proton;
we therefore do not consider $\gamma\gamma$~contributions in this
letter.

The LO amplitude for $q\bar q$~annihilation involves contributions containing two, one, or no
Z-boson propagators that may become resonant. 
At NLO, the same is true for $q\bar q$~amplitudes with EW loop
insertions and the corresponding amplitudes with real photonic
bremsstrahlung.  Since no couplings to W~bosons are involved at LO, we
can divide the EW corrections into separately gauge-independent
photonic and purely weak contributions.  By definition, the former
comprise all contributions with real photons and all loop and
counterterm diagrams with photons in the loop coupling to the external
fermions, while the latter are furnished by the remaining EW
corrections.  Actually the NLO EW corrections include contributions
from $q\gamma$, $\bar q\gamma$, and $\gamma\gamma$ channels as well,
but those contributions turn out to be phenomenologically unimportant.

Apart from the algebraic complexity, a major complication in the NLO
EW calculation arises from the appearance of resonances which require
at least a partial Dyson summation of the potentially resonant
self-energy corrections, a procedure that jeopardizes the gauge
invariance of the result if no particular care is taken.  We employ
the complex-mass scheme~\cite{Denner:1999gp,Denner:2005fg} which
provides a gauge-invariant solution to this problem at NLO by
replacing the real W- and Z-boson masses by complex quantities,
including also the corresponding complexification of EW couplings.  To
evaluate all one-loop integrals with complex W/Z~masses with
sufficient numerical stability in the four-body phase space, we apply
the library {\sc Collier}, which is mainly based on the results of
\citeres{Denner:2002ii,Denner:2005nn,Denner:2010tr} and briefly
described in \citere{Denner:2014gla}.

Infrared (soft and/or collinear) singularities in the real emission
amplitudes are extracted via dipole subtraction, as formulated in
\citeres{Dittmaier:2008md,Dittmaier:1999mb} for photon radiation.  The
infrared-singular contributions are alternatively treated in
dimensional or in mass regularization.  We have checked numerically
that the sum of all (virtual and real) corrections is infrared finite
and independent of the regularization scheme used.

We have performed two independent calculations of all contributions
and found results that are in mutual agreement within statistical
uncertainties of the final Monte Carlo phase-space integration.  
One calculation closely follows the strategy described in
\citeres{Denner:2005es,Denner:2005fg}, where NLO EW corrections to
$\Pep\Pem\to4\,$fermions via W-boson pairs were calculated,
in the loop part and builds on \citere{Billoni:2013aba}
in the real correction and the Monte Carlo integration.
The other calculation has been carried out with the program 
{\sc Recola}~\cite{Actis:2012qn,Uccirati:2014fda} 
facilitating the
automated generation of the NLO EW amplitudes, in combination with an
in-house Monte-Carlo generator.  Additional checks have been performed
employing the Mathematica package {\sc Pole}~\cite{Accomando:2005ra}.

\subsection{Input and event selection}

For the numerical analysis we consider the LHC running at
centre-of-mass (CM) energies of $7\TeV$,
$8\TeV$, $13\TeV$, and $14\TeV$ 
and choose the input parameters as follows.  The on-shell values for
the masses and widths of the gauge bosons,
\begin{equation}
\begin{array}[b]{rcl@{\quad}rcl}
  \MW^{\rm os} &=& 80.385  \GeV, & \GW^{\rm os} &=& 2.085 \GeV, \\
  \MZ^{\rm os} &=& 91.1876 \GeV, & \GZ^{\rm os} &=& 2.4952\GeV, 
\end{array}
\end{equation}
are first translated into the pole scheme according to
\begin{eqnarray}
&& M_V = M_V^{\rm os}/c_V, \quad \Gamma_V = \Gamma_V^{\rm os}/c_V, \nonumber\\
&& c_V=\sqrt{1+(\Gamma_V^{\rm os}/M_V^{\rm os})^2}, \quad V=\PW,\PZ,
\end{eqnarray}
and subsequently combined to complex mass parameters
$\mu_V^2=M_V^2-\ri M_V\Gamma_V$, as demanded by the complex-mass
scheme.  Since no other particles show up as resonances in our
calculation, their decay widths can be neglected and their masses
taken as on-shell parameters.
In detail, we set the Higgs-boson and the top-quark masses to 
\begin{equation}
\begin{array}[b]{rcl@{\quad}rcl}
\MH &=& 125\GeV,         & \Mt   &=& 173\GeV.
\end{array}
\end{equation}
We work 
in the $G_\mu$-scheme where    
the electromagnetic coupling $\alpha$ is derived from the Fermi constant
\begin{equation}
G_\mu= 1.16637\times 10^{-5} \GeV^{-2},
\end{equation}
according to
\begin{equation}
\alpha_{G_\mu} = \sqrt{2}G_\mu\MW^2\left(1-\MW^2/\MZ^2\right)/\pi.
\end{equation}
This choice absorbs the effect of 
the running of $\alpha$ to the
electroweak scale into the LO cross section and, thus, avoids mass
singularities in the charge renormalization. Moreover,
$\alpha_{G_\mu}$ partially accounts for the leading universal
renormalization effects originating from the $\rho$-parameter.  The
fine-structure constant
\begin{equation}
\alpha(0) = 1/137.035999679
\end{equation}
is only used as coupling parameter in the relative photonic
corrections, because those are strongly dominated by real photon
emission naturally coupling
with $\alpha(0)$.  The relative
genuine weak corrections, however, are parametrized with
$\alpha_{G_\mu}$.

The renormalization and factorization scales, $\mu_{\mathrm{ren}}$ and
$\mu_{\mathrm{fact}}$, are set equal to the pole mass of the Z~boson,
$\mu_{\mathrm{ren}}=\mu_{\mathrm{fact}}=\MZ$.  Since we focus on EW
corrections, we consistently employ the set
NNPDF2.3QED~\cite{Ball:2013hta} of parton distribution functions
(PDFs), which are the only up-to-date PDFs including QED corrections.

In the event selection, we apply a set of phase-space cuts 
that is optimized for Higgs studies, 
inspired by the CMS and ATLAS analyses~\cite{Chatrchyan:2013mxa,Aad:2014eva}. 
For each lepton $\ell_i$, we exclude too low transverse momentum and too large rapidity demanding
\begin{equation}
\label{eq:inc-cuts1}
 p_\rT(\ell_i) > 6\GeV, \quad |y(\ell_i)| < 2.5,
\end{equation}
and any pair of charged leptons is required to be well separated in the rapidity--azimuthal-angle plane, 
\begin{equation}
\label{eq:inc-cuts2}
\Delta R(\ell_i,\ell_j) = \sqrt{(y_i-y_j)^2+(\phi_i-\phi_j)^2}>0.2.
\end{equation}
Photons are recombined with the closest $\ell_i$ if 
\begin{equation}
\label{eq:recombi}
\Delta R(\gamma,\ell_i) < 0.2.
\end{equation}

To 
these basic selection criteria we add cuts on the invariant masses
$M_{\ell_i^+\ell_i^-}$ of the $\ell_i^+\ell_i^-$~pairs,
\begin{eqnarray}
\label{eq:hcuts1}
  40\GeV & < M_{\ell_1^+\ell_1^-} < & 120\GeV, \nonumber\\
  12\GeV & < M_{\ell_2^+\ell_2^-} < & 120\GeV,
 \end{eqnarray}
with $\ell_1^+\ell_1^-$ (${\ell_2^+\ell_2^-}$) referring to the $\ell^+\ell^-$ pair that is 
closer to (further away from) the nominal mass of the Z~boson.  
Moreover, we impose a cut on the invariant mass $M_{4\ell}$ of the four-lepton system,
\begin{equation}
\label{eq:hcuts2}
  M_{4\ell} > 100\GeV.
\end{equation}%

\subsection{Numerical results}

In the following we discuss the LO cross section $\sigma_{\bar q
  q}^{\LO}$ for $\ppmmee$ and the corresponding 
full EW and purely weak
relative corrections 
$\delta_{\bar q q}^{\EW}=\delta_{\bar q q}^{\mathrm{photonic+weak}}$ 
and $\delta_{\bar q q}^{\weak}$, 
which are normalized to $\sigma_{\bar q q}^{\LO}$.  The
label ${\bar q q}$ indicates that only quark--antiquark annihilation
channels are taken into account.  The relative corrections
$\delta_{\bar q q}^{\EW}$ and $\delta_{\bar q q}^{\weak}$ are rather
insensitive to the PDF set and, thus, can be used to promote QCD-based
cross sections to state-of-the-art predictions via reweighting.

Table~\ref{tab:xsec} shows $\sigma_{\bar q q}^{\LO}$, $\delta_{\bar q
  q}^{\EW}$, 
and $\delta_{\bar q q}^{\weak}$ for various LHC energies.
\begin{table}
\begin{center}
\begin{tabular}
{|c|ccc|}
\hline
$\sqrt{s}$~[TeV] & $\sigma_{\bar q q}^{\LO}$~[fb] &  $\delta_{\bar q q}^{\EW}~[\%]$& $\delta_{\bar q q}^{\weak}~[\%]$ 
\\
\hline
$7$ & $\phantom{1}7.3293(4)$ & $-3.4$& $-3.3$ 
\\
$8$ & $\phantom{1}8.4704(2)$ & $-3.5$& $-3.4$
\\
$13$ & $13.8598(3)$  & $-3.6$ &$-3.6$ 
\\
$14$ & $14.8943(8)$ & $-3.6$ &$-3.6$ 
\\
\hline
\end{tabular}
\end{center}
\caption{LO cross section $\sigma_{\bar q q}^{\LO}$ for $\ppmmee$ for various LHC energies
and corresponding EW 
($\delta_{\bar q q}^{\EW}$)
and purely weak relative corrections
($\delta_{\bar q q}^{\weak}$).}
\label{tab:xsec}
\end{table}
The integrated cross sections are, of course, dominated by resonant
Z-boson pair production, i.e.\ by partonic CM energies 
$\sqrt{\hat s}>2\MZ$. Accordingly, the EW corrections largely resemble
the size of the known corrections to on-shell
ZZ~production~\cite{Bierweiler:2013dja,Baglio:2013toa}, which amount
to about $\sim -4.5\%$.  The remaining $\sim1\%$ can be attributed to
the EW corrections to non-resonant contributions and the
acceptance effects on leptonic Z-boson decays.
Although the corrections to Z-boson decays are known to be small in an
inclusive setup (at the level of few per mille), in the presence of
the applied acceptance cuts they are enhanced to few percent, 
mainly as a result of the
sensitivity to final-state radiation (FSR).  In summary, our results
confirm that the NLO EW corrections to the cross section of Z-pair
production (including Z~decays and off-shell effects) are at the
$5\%$~level at the LHC and, thus, have to be taken into account in the
confrontation of data with theory.  The major part of the EW
corrections is due to genuine weak effects, while photonic corrections
remain below
the $1\%$ level.

\begin{figure}
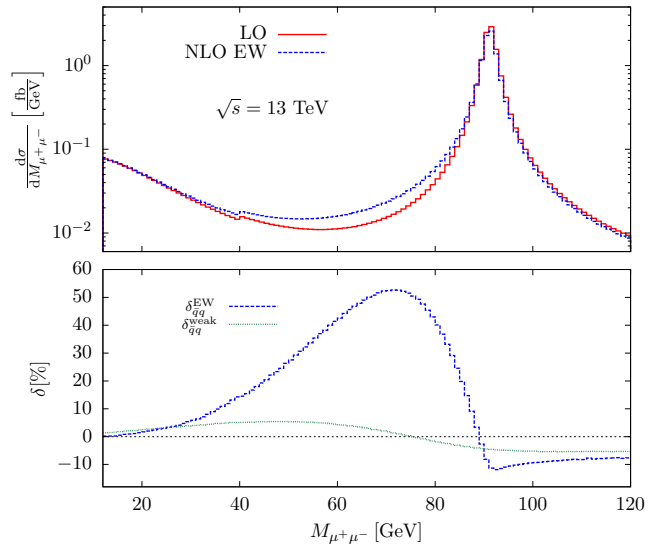

 \includegraphics[width=.47\textwidth]{{{HiggsBG.13TeV.6GeVptcut.invmass.mu+mu-}}}
\vspace*{-1em}
\caption{Invariant-mass distribution of the $\mu^+\mu^-$~system in $\Pp\Pp\to\mu^+\mu^-\Pep\Pem+X$
  including NLO EW corrections (upper panel), and relative EW and
  purely weak corrections at NLO (lower panel).}
\label{fig:Mmm}
\end{figure}

We turn to differential distributions at $13\TeV$, 
focussing on kinematical
variables that are particularly sensitive to the off-shellness of the
intermediate Z~bosons, i.e.\ to distributions that are not accessible
by previous calculations based on on-shell Z~bosons.
Figure~\ref{fig:Mmm} shows the invariant-mass distribution of the
$\mu^+\mu^-$~system and the relative full EW and weak corrections.
Note that the distribution is dominated by resonant $\Pep\Pem$~pairs
throughout ($M_{\Pep\Pem}\sim\MZ$). In the vicinity of the Z-boson
resonance ($M_{\mu^+\mu^-}\sim\MZ$), the weak corrections, thus,
mainly result from the following two contributions of different
origin: First, there is a constant offset of $\sim-5\%$ stemming
mainly from weak corrections to the dominant
$\Pp\Pp\to\PZ({\to}\Pep\Pem)\PZ^*$ production with a resonance in
$M_{\mu^+\mu^-}$.  Second, there are the weak corrections to the
interference of the resonant and non-resonant contributions to the
amplitude. These second contributions are proportional to
($M_{\mu^+\mu^-}^2-\MZ^2$) and thus change sign at the Z~resonance.
The pronounced shapes of the EW and weak
corrections in fact largely resemble the structures known from
single-Z production (see, e.g., Fig.~12 of \citere{Dittmaier:2009cr}),
with the large radiative tail for $M_{\mu^+\mu^-}\lsim\MZ$ originating
from FSR.  While FSR effects can be reproduced by photonic parton
showers quite well, the genuine weak corrections cannot be
approximated easily.  As in the case of single-Z production, the weak
corrections exhibit a sign change near the resonance (shifted to
smaller $M_{\mu^+\mu^-}$ in \reffi{fig:Mmm} because of the negative
offset mentioned above).
Far below the Z-boson resonance the relative EW
corrections do not show large variations.  This fact 
is
interesting in view of the Higgs-boson signal resulting from
$\Pp\Pp(\Pg\Pg)\to\PH\to\mu^+\mu^-\Pep\Pem+X$ (not shown here), whose
$M_{\mu^+\mu^-}$ distribution shows a shoulder for
$M_{\mu^+\mu^-}\lsim\MH-\MZ\approx34\GeV$ sensitive 
to the quantum numbers of the Higgs boson \cite{Choi:2002jk}.

\begin{figure}
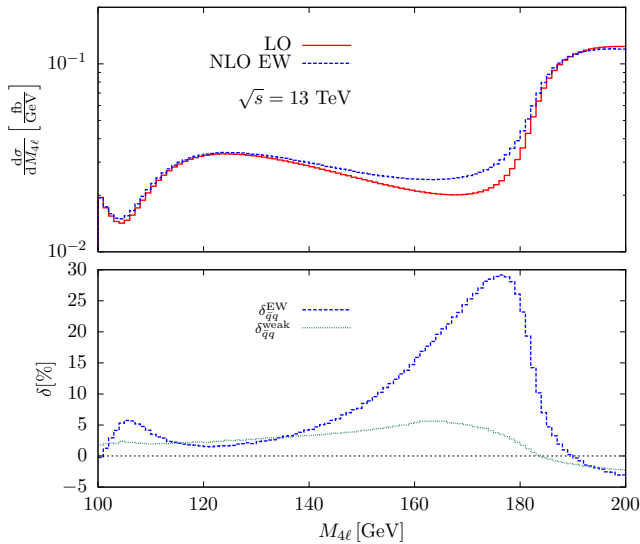

\includegraphics[width=.47\textwidth]{{{HiggsBG.13TeV.6GeVptcut.invmass.mu+mu-e+e-}}}
\vspace*{-1em}
\caption{Four-lepton invariant-mass distribution in $\Pp\Pp\to\mu^+\mu^-\Pep\Pem+X$
including NLO EW corrections (upper panel), and relative EW and purely weak corrections at NLO 
(lower panel).}
\label{fig:Mmmee}
\end{figure}

In \reffi{fig:Mmmee} we show 
the invariant-mass distribution of the full
four-lepton system, which features 
the Higgs resonance from $\Pg\Pg$~fusion at
$M_{4\ell}\sim\MH\approx125\GeV$ (not included here). 
The steep shoulder at the Z-pair threshold at
$M_{4\ell}=2\MZ\approx182\GeV$ creates a radiative tail at smaller
invariant masses, similar to the case of the $M_{\mu^+\mu^-}$
distribution, since $M_{4\ell}$ can be strongly decreased by FSR
effects.  A similar effect, though reduced, is observed below the
second shoulder near $M_{4\ell}=110\GeV$, which is a result of the
$p_\rT$ and invariant-mass cuts \refeq{eq:inc-cuts1} and
\refeq{eq:hcuts1}.  In the region of the Higgs-boson resonance the EW
corrections are at the level of a few percent.  While photonic
corrections might again be well approximated by parton showers, this
does not apply to the weak corrections.  Interestingly, the weak
corrections change their size from $-3\%$ to about $+6\%$
when
$M_{4\ell}$ drops below the Z-pair threshold.  The sign change can be
understood from the fact that below the ZZ~threshold one of the two
Z~bosons is forced to be far off shell. For the corresponding
$\ell^+\ell^-$ pair, this means that $M_{\ell^+\ell^-}$
drops below $\MZ$, so that the weak corrections turn positive, as can
be seen from \reffi{fig:Mmm}.
The sign change of the weak corrections near the ZZ~threshold is quite
interesting phenomenologically, since it renders their inclusion via a
global rescaling factor impossible.  Globally reducing differential
cross sections by $3.6\%$, as deduced from the integrated cross
section, would have the opposite effect on the $M_{4\ell}$
distribution near the Higgs signal as the true weak correction.

\begin{figure}
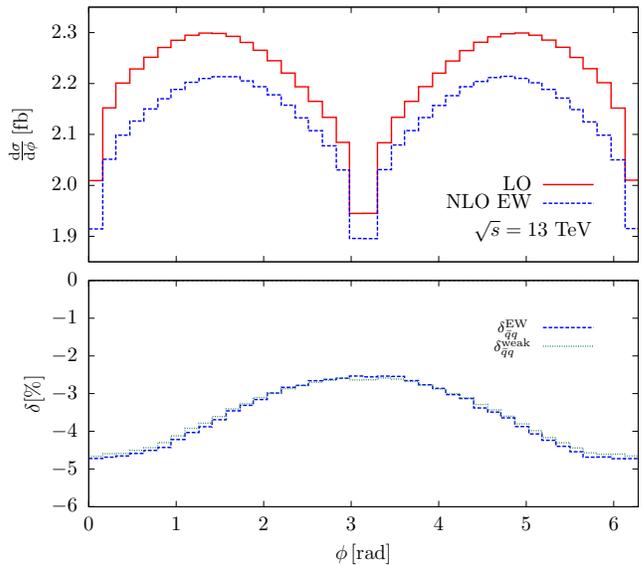

\includegraphics[width=.47\textwidth]{{{HiggsBG.13TeV.6GeVptcut.ZZdecayAngle}}}
\vspace*{-1em}
\caption{Distribution in the angle $\phi$ between the two $\PZ$-boson decay planes
in $\Pp\Pp\to\mu^+\mu^-\Pep\Pem+X$
including NLO EW corrections (upper panel), and relative EW and purely weak corrections at NLO 
(lower panel).}
\label{fig:phi}
\end{figure}

Finally, in \reffi{fig:phi} we show the distribution in the angle
$\phi$ between the two $\PZ$-boson decay planes, which are each
spanned by the two lepton momenta of the respective
$\ell^+\ell^-$~pair \cite{Bredenstein:2006rh}.
The distribution is sensitive to possible deviations of the
Higgs-boson coupling structure from the Standard Model prediction, so
that any distortion of the distribution induced by higher-order
corrections, if not properly taken into account, could mimick
non-standard effects. Figure~\ref{fig:phi} reveals a distortion by
about $2\%$ 
due to weak loop effects. The contribution of
photonic corrections is negligible in our setup, 
similar to their
contribution to the integrated cross section.  This is due to the fact
that photonic corrections mainly influence the absolute size of the
lepton momenta via collinear FSR, but not the directions of the
leptons.

In summary, the NLO EW corrections to four-lepton production consist
of photonic and purely weak contributions displaying
rather different features.  Photonic corrections can grow very large,
to several tens of percent, in particular in distributions where
resonances and kinematic shoulders lead to radiative tails.  While
those corrections might be well approximated with parton showers, this
is not the case for the remaining weak corrections, which are
typically of the size of $5\%$ and, thus, non-negligible.  The weak
corrections, in particular, distort distributions that are important
in Higgs-boson analyses. In the four-lepton invariant mass, even the
signs of the weak corrections in the Higgs signal region and the region
of resonant Z-boson pairs are different.

\vspace{1em}

B.J.\ gratefully acknowledges L.~Salfelder for useful discussions.
The work of S.D.\ is supported by the Research Training Group GRK~2044
of the German Science Foundation (DFG). A.D.\ and B.B.\ acknowledge
support by the DFG under reference number DE~623/2-1.  The work of
L.H.\ was supported by the grants FPA2013-46570-C2-1-P and
2014-SGR-104, and partially by the Spanish MINECO under the project
MDM-2014-0369 of ICCUB (Unidad de Excelencia `María de Maeztu').  The
work of B.J.\ is supported in part by the Institutional Strategy of
the University of T\"ubingen (DFG, ZUK~63) and in part by the German
Federal Ministry for Education and Research (BMBF) under contract
number 05H2015.

\bibliography{ppzz4l-letter}

\end{document}